\documentclass[aps,twocolumn,amssymb,prl,superscriptaddress,unsortedaddress]{revtex4}
\usepackage{epsf}
\usepackage{graphicx}
\usepackage{color}

\begin{document}

\title{Anomalous Fermi pockets on Hund's metal surface of Sr$_2$RuO$_4$\\
induced by the correlation-enhanced spin-orbit coupling}

\author{Takeshi~Kondo}
\altaffiliation{Corresponding author: kondo1215@issp.u-tokyo.ac.jp}
\affiliation{ISSP, The University of Tokyo, Kashiwa, Chiba 277-8581, Japan}
\affiliation{Trans-scale Quantum Science Institute, The University of Tokyo, Tokyo 113-0033, Japan}

\author{Masayuki~Ochi}
\affiliation{Department of Physics, Osaka University, Toyonaka, Osaka 560-0043, Japan}
\affiliation{Forefront Research Center, Osaka University, Toyonaka, Osaka 560-0043, Japan}

\author{Shuntaro~Akebi}
\affiliation{ISSP, The University of Tokyo, Kashiwa, Chiba 277-8581, Japan}

\author{Yuyang~Dong}
\affiliation{ISSP, The University of Tokyo, Kashiwa, Chiba 277-8581, Japan}

\author{Haruka~Taniguchi}
 \affiliation{Department of Applied Physics, Nagoya University, Nagoya 464-8603, Japan}
\affiliation{Department of Physics, Kyoto University, Kyoto 606-8502, Japan}

\author{Yoshiteru~Maeno}
\affiliation{Department of Physics, Kyoto University, Kyoto 606-8502, Japan}
\affiliation{Toyota Riken - Kyoto University Research Center (TRiKUC), Kyoto 606-8501, Japan}

\author{Shik~Shin} 
\affiliation{ISSP, The University of Tokyo, Kashiwa, Chiba 277-8581, Japan}
\affiliation{Office of University Professor, The University of Tokyo, Chiba 277-8581, Japan}

\date{\today}
\begin{abstract}
{The electronic structure of the topmost layer in Sr$_2$RuO$_4$ in the close vicinity of the Fermi level is investigated by angle-resolved photoemission spectroscopy (ARPES) with a 7-eV laser. We find that the spin-orbit coupling (SOC) predicted as 100 meV by the density functional theory (DFT) calculations is enormously enhanced in a real material up to 250 meV, even more than that of bulk state (200 meV), by the electron-correlation effect increased by the octahedral rotation in the crystal structure.
 This causes the formation of highly orbital-mixing small Fermi pockets and reasonably explains why the orbital-selective Mott transition (OSMT) is not realized in perovskite oxides with crystal distortion. Interestingly, Hund’s metal feature allows the quasiparticle generation only near $E_{\rm F}$, restricting the spectral gap opening derived by band hybridization within an extremely small binding energy ($<$ 10 meV). Furthermore, it causes coherent-incoherent crossover, making the Fermi pockets disappear at elevated temperatures. The anomalous Fermi pockets are characterized by the dichotomy of the orbital-isolating Hund’s coupling and the orbital-mixing SOC, which is key to understanding the nature of Sr$_2$RuO$_4$.
}
\end{abstract}

\maketitle

The effect of spin-orbit coupling (SOC) on the electronic properties of matter is one of the central topics in modern condensed matter physics. In particular, much attention has been recently given to the SOC effect of strongly
 correlated systems ~\cite{Borisenko,Balents,Damascelli_IronSC,Taylor,Johnson,Stadler,Day,Vale,HiroshiShinaoka,Kim2018hj,Arita,Zhang2016bz,Tamai2019ef,Damascelli2014,Iwasawa,Haverkort,DJKim_PRL,LukeJSandilands,Zwartsenberg,KimAaram,Dhital,Das,ZhouSen,Cui,Watanabe,Nie,Martins}. Transition metal perovskite oxides provide an excellent platform to study the cooperative effects of electron correlation and SOC on the band structure~\cite{Arita,Zhang2016bz,Kim2018hj,Tamai2019ef,Damascelli2014,Iwasawa,Haverkort,DJKim_PRL,LukeJSandilands,Zwartsenberg,KimAaram,Dhital,Das,ZhouSen,Cui,Watanabe,Nie,Martins}. Theory predicts that the electron correlation enhances SOC, increasing the energy splitting of the bands derived by multi $d$-orbitals~\cite{Kim2018hj,Zhang2016bz}. In contrast, the band narrowing due to the correlation effect reduces the band splitting, canceling the effect by the enhanced SOC. As a result, the SOC-induced band splitting may become comparable in energy with that obtained by correlation-free DFT calculations. The validity of this argument has been confirmed for the bulk state of Sr$_2$RuO$_4$ by ARPES measurements~\cite{Tamai2019ef}.

Another intriguing and common aspect of perovskite oxides is the octahedral rotation in the crystals~\cite{Rotation_PRB,Sr3Ir2O7_rotation,Sr3Ru2O7_rotation,Sr2IrO4_rotation,CadopeRotation,Braden4,Moore_SurfacePhase,STM,Ko_DFT}. This could stimulate orbital mixing by band folding and enhances the correlation effect by reducing hopping integral. For example, the diverse electronic properties of Ca$_{2-x}$Sr$_x$RuO$_4$ (4$d$-system) are controlled by Ca substitution that induces a rotation as well as the tilt of RuO$_6$-octahedra~\cite{CadopeRotation}. In this system, the orbital-selective Mott transition (OSMT) has been theoretically proposed~\cite{AnisimovEPB}; however, it has not been realized in experiments, calling the OSMT scenario into question~\cite{Liebsch_2003,LiebschPRL2003,LiebschPRB,WangYangPRL,ShimoyamadaPRL}. 
 In Sr$_2$IrO$_4$ (5$d$-system), the IrO$_6$-rotation naturally occurs~\cite{Ir_rotation}, and a Fermi arc similar to that of underdoped cuprates was observed in the electron-doped surface~\cite{Kim_arc1,Kim_arc2}.
 The analogy to cuprates (3$d$-system) is, however, still debated since the intrinsic structure could be Fermi pockets derived from the IrO$_6$-rotation and/or the antiferromagnetic order rather than the Fermi arc \cite{Sr2IrO4_Baumberger,Sr2RhO4_Kim,Sr2RhO4_Baumberger,Damascelli_FS,Shen_surface,Damascelli_Progression}.
 As apparent from these, uncovering how the octahedral rotation modifies the electron correlation and SOC effects on the band structure is crucial for establishing the physics of multi-orbital correlated systems. 
 
Sr$_2$RuO$_4$ has a unique surface layer of RuO$_6$-octahedrons rotated by $\sim8^\circ$~\cite{Rotation_PRB,STM,Moore_SurfacePhase,Damascelli_Progression},
which contrasts the underlying bulk with no such distortion. The surface does not have the tilt of octahedra nor magnetic order, so it allows one to investigate the effect only of the octahedral rotation on the band structure. 
In particular, the availability of comparing the surface state with the underlying bulk system without the octahedral rotation is quite advantageous for precisely pinning down the rotation effect. The ARPES signals of the Sr$_2$RuO$_4$ surface are usually contaminated by the bulk signals, which prevents a detailed examination of the surface bands. This difficulty can, however, be solved by using low-energy photons in ARPES, which observe only the topmost layer of the Sr$_2$RuO$_4$ crystals~\cite{KondoPRL}. 

The Sr$_2$RuO$_4$ surface is further known to be in Hund’s metal state, where Hund’s coupling $J$ plays a crucial role in the strong correlation~\cite{KondoPRL,HundMetal1,HundMetal_NatNano,HundMetal_resisitivity,Hund_Seebeck,HundMetal2,Crossover,HundMetal_PRB}. Hund’s coupling effectively isolates the orbital character of each band, in stark contrast to SOC, which facilitates orbital mixing. Hence, observation of the Sr$_2$RuO$_4$ surface by a low-energy laser ARPES enables one to unveil electronic properties intricated by the mutual relation among strong electron correlation, spin-orbit coupling, octahedral rotation, and Hund’s coupling. These are rather general in multi-orbital correlated systems, thus quite important, yet have not been revealed to date.

In this letter, we use high energy and momentum resolutions laser-ARPES ($h\nu=7$eV) and investigate the electronic states in close vicinity of the Fermi level on the reconstructed Sr$_2$RuO$_4$ surface. We unveil the anomalous nature of Fermi pockets originating from the strong correlation effects tied with the dichotomic features of Hund’s coupling and spin-orbit coupling that tend to separate and mix orbitals, respectively.

Single crystals of Sr$_2$RuO$_4$ were grown by the floating-zone technique~\cite{MAO20001813}. 
ARPES measurements were performed for the (001) cleaving surface with a Scienta R4000 analyzer equipped with a 7~eV laser at The Institute for Solid State Physics (ISSP), The University of Tokyo. The energy resolution was $\sim2$ meV, and the lowest measured temperature was 5K. Details of band calculations are described in Supplemental Material~\cite{SM}. 

\begin{figure}
\includegraphics[width=3.5in]{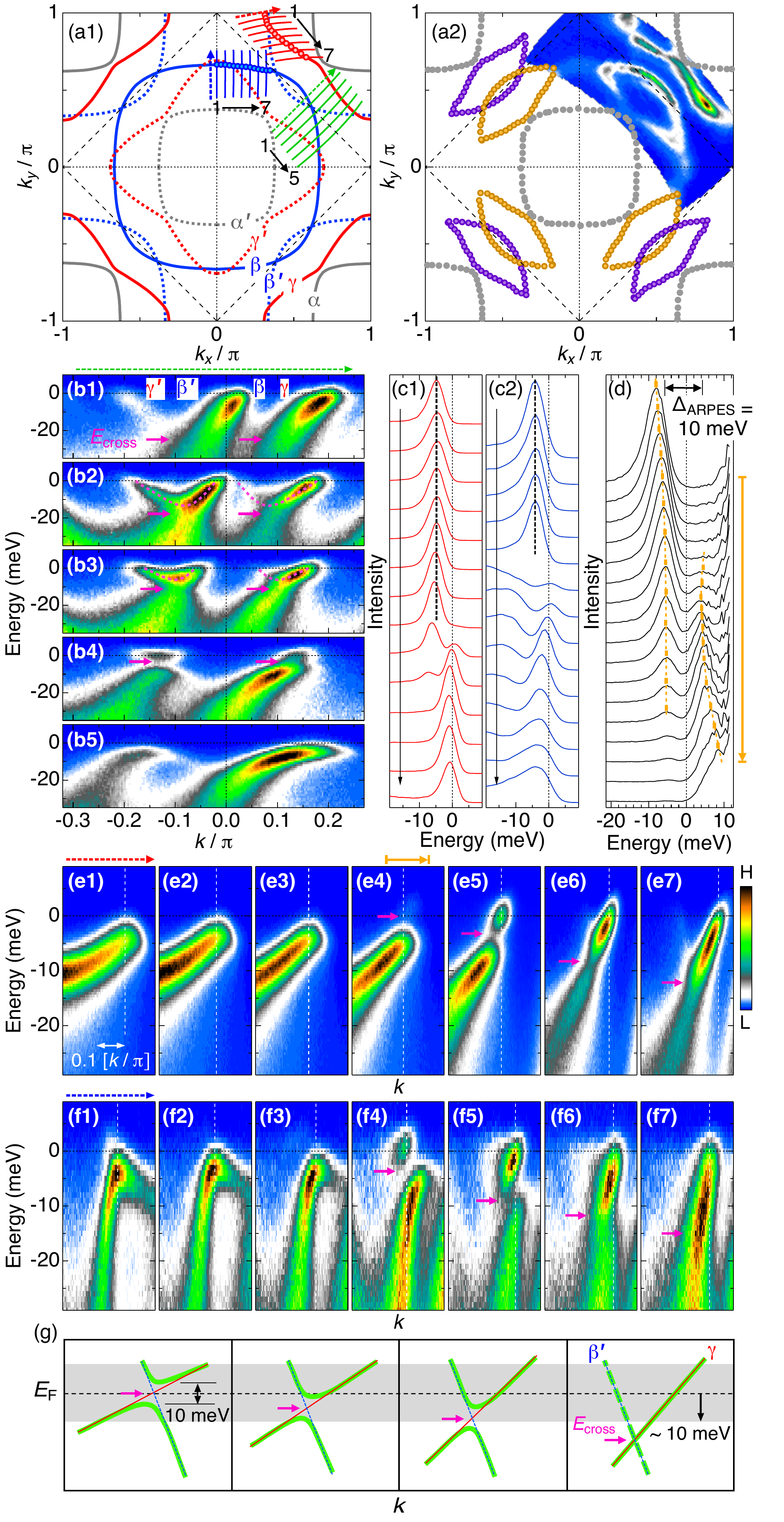}
\caption{Fermi pockets. (a1) Schematic FSs of Sr$_2$RuO$_4$.
The main FS (solid curves) are folded (dotted curves) about the BZ (dashed lines) reduced by octahedral rotation. 
(a2) FS mapping for Fermi pockets and determined FSs. $\alpha$-FS was determined from the previous data~\cite{KondoPRL}.
(b1-b5) ARPES dispersions along green momentum cut 1 to 5 in (a1). 
(c1,c2) EDCs at $k$s marked by white dotted lines in (e1-e7) and (f1-f7), respectively. Peak shifts due to a hybridization gap of 4 to 5 meV are indicated by dashed lines. (d) EDCs along the orange arrow in (e4) divided by the Fermi function at the measured temperature (20 K). 
(e1-e7, f1-f7) ARPES dispersions along red and blue momentum cut 1 to 7 in (a1), respectively. The data other than (d) were taken at 5 K.
(g) Schematic for (e4-e7) showing band crossing between $\gamma$- and $\beta'$-bands. A hybridization gap opens only when $E_{\rm cross}$ is less than $\sim$10 meV in the binding energy.}  
\label{fig1}
\end{figure}

The Fermi surface (FS) of the Sr$_2$RuO$_4$ surface is complexed by the folding back of bands about the zone boundaries reduced by the RuO$_6$ rotation~\cite{STM,Moore_SurfacePhase} [dashed black lines in Fig.~1(a1)]. In Fig.~1(a1), the schematic FSs for $\alpha$- and $\beta$-bands derived from the $d_{xz/yz}$-orbitals, for $\gamma$-band from the $d_{xy}$-orbital, and also for their folded bands ($\alpha'$-, $\beta'$-, and $\gamma'$-bands) are illustrated (gray, blue, and red curves, respectively). The main $\gamma$-FS before the band folding is hole-type centered at ($\pi, \pi$), which differs from the electron-type centered at (0, 0) in the bulk state. This variation of FS-topology is caused by a relative energy shift of the saddle point at ($\pi$,0) due to the structural distortion on the surface~\cite{Ko_DFT}. 

Figures~1(b1)-1(b5) show the ARPES dispersions along green momentum cuts 1 to 5 in Fig.~1(a1) close to the Fermi energy ($E_{\rm F}$). The $\gamma'$- and $\beta$-bands (or, the $\gamma$- and  $\beta'$-bands) mutually cross at $k_{\rm cross}$ [see inset of Fig.~3(b)], and the crossing energy ($E_{\rm cross}$) marked by magenta arrows shifts with different momentum cuts. We find that these bands independently disperse with no clear hybridization when $E_{\rm cross}$ is far from $E_{\rm F}$ [Fig.~1(b1)]. As $E_{\rm cross}$ gets closer to $E_{\rm F}$, the $\gamma'$- and $\beta$-bands as well as the $\gamma$- and $\beta'$-bands each hybridize to open a gap~\cite{KingSurfacePRL,Peter_Advanced}, generating two parabolic bands [dotted curves in Fig.~1(b2) and 1(b3)]. The parabolas become smaller and eventually disappear into the unoccupied side [Fig.~1(b4)], leaving spectral tails on the occupied side. 

We reveal here that a gap opens at $E_{\rm F}$ over a wide momentum region, yielding small Fermi pockets. 
Figures~1(e1)-1(e7) plot the ARPES dispersions along red momentum cuts in Fig.~1(a1) accessing up to the zone edge. 
When $E_{\rm cross}$ (marked by magenta arrows) gets close to $E_{\rm F}$, a hybridization gap opens [Fig.~1(e6)]. 
The gap position gradually shifts up toward $E_{\rm F}$ [Fig.~1(e5)] as momentum approaches the zone edge. Once reaching $E_{\rm F}$ [Fig.~1(e4)], the gap stays there up to the zone edge [Fig.~1(e1)]. 
This behavior is schematically illustrated in Fig.~1(g). 
 Note here that in the ARPES data, the spectral intensity of the $\beta$-band folded due to the reduced BZ is weak. 
 In the reduced BZ, the momentum region marked by blue arrows in Fig.~1(a1) is equivalent to that marked by red arrows. Indeed, a similar gap behavior is observed [Figs.~1(f1)-1(f7)], while the spectral intensity of the folded $\alpha$-band is weak in this case. 
 
 The gap opening at $E_{\rm F}$ is better demonstrated in Fig.~1(c1) and Fig.~1(c2) by extracting energy distribution curves (EDCs) at $k_F$s [circles in Fig.~1(a1)]. The momentum area with a gap opening is wide, and consequently, two small pockets are formed around ($\pi/2, \pi/2$) [Fig.~1(a2)]. To reveal the band dispersion slightly above $E_{\rm F}$, we raise the sample temperature to 20 K, and plot in Fig.~1(d)  the Fermi-Dirac distribution function (FD) divided EDCs obtained along the orange arrow in Fig.~1(e4). The band gap between the parabolic conduction and valence bands ($\Delta_{\rm ARPES}$) is estimated as $\sim$10 meV~\cite{KingSurfacePRL,Peter_Advanced}. The gap magnitude is almost constant at different momenta as long as $E_{\rm cross}$ is close to $E_{\rm F}$ (Supplemental Fig. S2~\cite{SM}).

\begin{figure}
\includegraphics[width=3.4in]{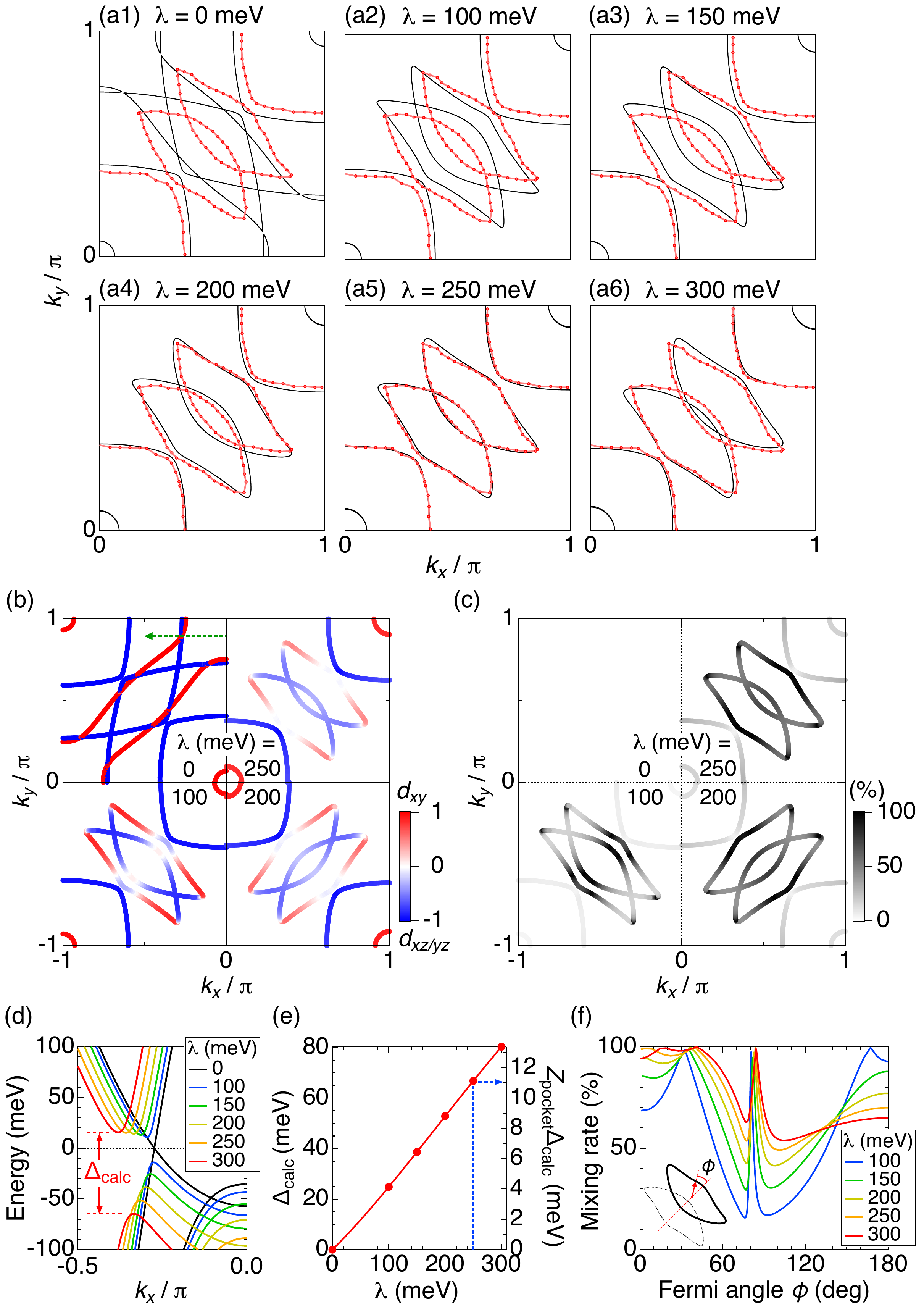}
\caption{Band calculations with enhanced SOC reproducing ARPES data. 
(a1-a6) Comparison of FSs by ARPES (red dots) and calculations (black curves) with different SOC strengths ($\lambda$). (b) FS colored by red ($d_{xy}$) and blue ($d_{xz/yz}$) representing dominant orbital. The color gets white when the mixing ratio is high. 
(c) The mixing rate is represented by a gray scale. Black corresponds to a 100$\%$ mixing, at which FS is evenly contributed by $d_{xy}$ and $d_{xz/yz}$ orbitals. 
(d) Band dispersions along the green dashed arrow in (b) calculated for 
several different $\lambda$. (e) Calculated hybridization gap ($\Delta_{\rm calc}$) plotted against $\lambda$. As an example, $\Delta_{calc}$ for $\lambda$ = 300 meV is represented by the dimension arrow in (d). The right axis indicates $\Delta_{calc}$ renormalized by $Z_{\rm pocket}$$\equiv$$\sqrt{Z_{\beta}Z_{\gamma}}$; 11 meV is obtained at $\lambda$ = 250 meV. (f) The mixing rate plotted against the Fermi angle $\phi$ defined in the inset.
}  
\label{fig1}
\end{figure}

To better understand the Fermi pockets and hybridization gap, we perform the DFT calculations for Sr$_2$RuO$_4$ under the RuO$_6$ rotation of $\sim8^\circ$. The Fermi surfaces (FSs) obtained with and without SOC are shown in Fig.~2(a1) and 2(a2), respectively, together with ARPES results (red lines and circles). Here, the SOC strength (${\lambda}$) is estimated to be 100 meV by DFT calculations ($\lambda_{\rm DFT}$). The Fermi pockets are obtained only when SOC is included. Note that a hybridization gap opens at the momenta of intersection between the two Fermi pockets, but it is very small. Importantly, we find that the Fermi pockets of DFT are profoundly mismatched with those of ARPES: the overlapped area of the two pockets is much smaller in the data. 
By increasing ${\lambda}$ [Fig.~2(a3-a6)], the mismatch is mitigated, and eventually, a good matching with ARPES results is obtained at ${\lambda}$ of 250 meV. The mismatch gets pronounced again with increasing ${\lambda}$ further by exhibiting   
the overlapped area to be too small. 

According to theory~\cite{Zhang2016bz,Kim2018hj}, the SOC strength is effectively enhanced by electron correlations: ${\lambda}^{\rm eff}={\lambda}_{\rm DFT}+{\Delta}{\lambda}$. Here, ${\Delta}{\lambda}$ is the correlation-enhanced value.
Our results [Fig.~2(a5)] determines ${\Delta}{\lambda}$ = 150meV for the reconstructed surface state. 
Interestingly, the ${\Delta}{\lambda}$ is about 1.5 times higher than that in the bulk state (100 meV~\cite{Tamai2019ef}), indicating that the surface state of Sr$_2$RuO$_4$ is more strongly correlated than the underlying bulk state, most likely due to the octahedral rotation. 
This is, indeed, reflected in the difference of the band renormalization factor ($m^*/m_{\rm calc}=v_{\rm calc}/v_F=1/Z$) between the two states. Here, $m^*$ ($v_F$) and $m_{\rm calc}$ ($v_{\rm calc}$) are masses (velocities) obtained by experiments and correlation-free calculations, respectively, and $Z$ is quasiparticle residue. It is reported that $1/Z$ = 5.9, 4.2, and 8.6 for $\alpha$, $\beta$, and $\gamma$ bands, respectively, for the surface states~\cite{KondoPRL}, which are much larger than those (3.0, 3.5, and 5.5, respectively~\cite{Maeno}) for the bulk state. $1/Z$ of $d_{xz/yz}$ and $d_{xy}$ ($1/Z_{xz/yz} \equiv 1/\sqrt{Z_{\alpha}Z_{\beta}}$ and $1/Z_{xy} \equiv 1/Z_{\gamma}$) are each 4.98 and 8.6 for the surface state and each 3.24 and 5.5 for the bulk state. Therefore, the bands derived by $d_{xz/yz}$ and $d_{xy}$ are both $\sim$1.5 times more largely renormalized in the surface state than in the bulk state. Notably, this is consistent with the ratio of the correlation-enhanced value of ${\Delta}{\lambda}$ between these two states.

\begin{figure}
\includegraphics[width=3.4in]{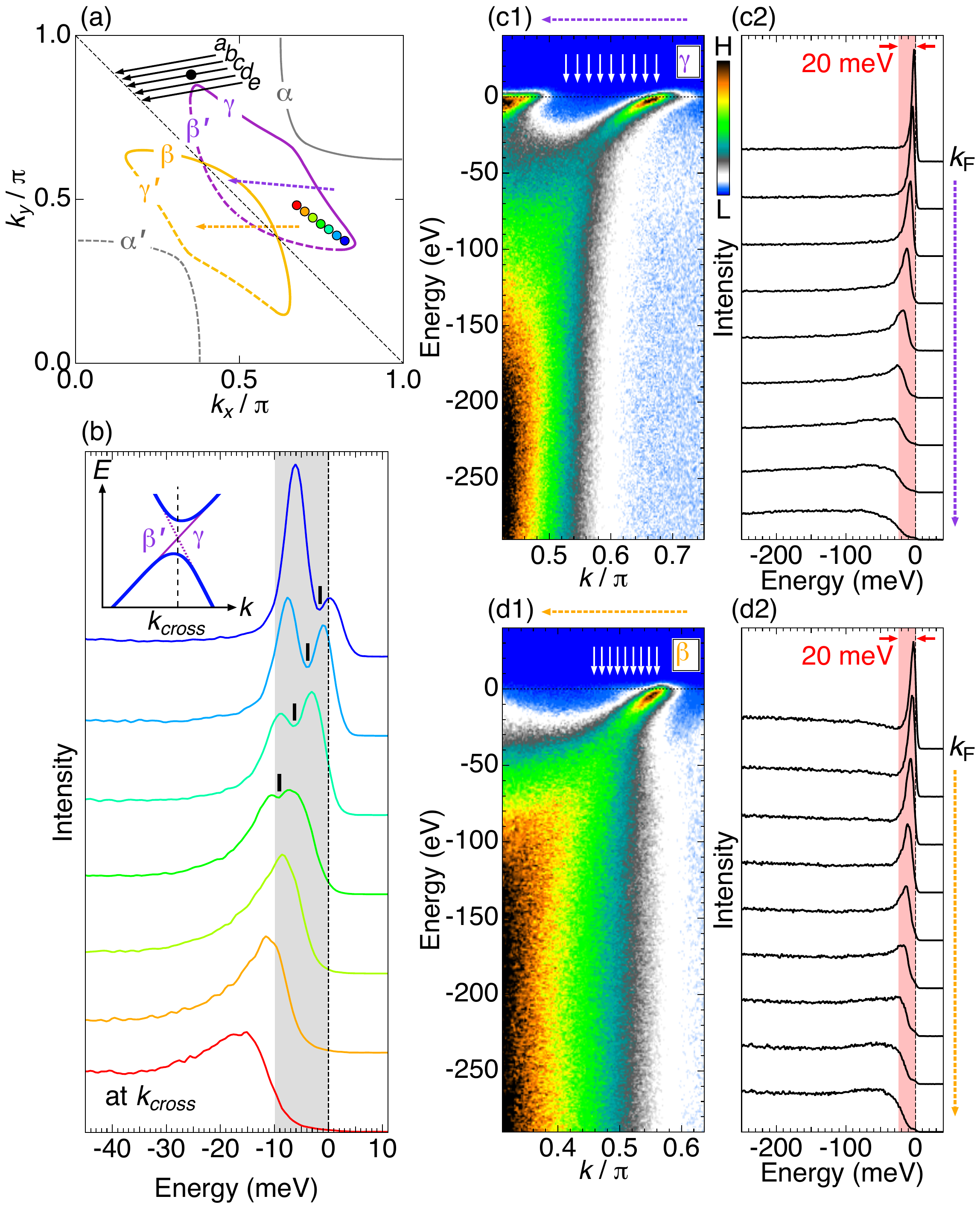}
\caption{Collapse of hybridization gap with binding energy. (a) FS on the Sr$_2$RuO$_4$ surface. The main ($\alpha$, $\beta$, and $\gamma$) and folded ($\alpha'$, $\beta'$ and $\gamma'$) parts of FS are each represented by solid and dashed curves. (b) EDCs at $k_{\rm cross}$ at which $\gamma$- and $\beta'$-band cross, as depicted in the inset. The energy window of 10 meV is hatched by gray. The spectral dip due to hybridization gap is marked by a bar.
(c1,c2) ARPES dispersion along the purple arrow in (a) mainly capturing $\gamma$ band and ECDs extracted at $kx$ marked by white arrows in (c1), respectively. The energy window of 20 meV is hatched by red.
(d1,d2) Similar data to (c1,c2), but along the yellow arrow in (a) mainly capturing $\beta$ band.   
}  
\label{fig1}
\end{figure}

\begin{figure}
\includegraphics[width=3.4in]{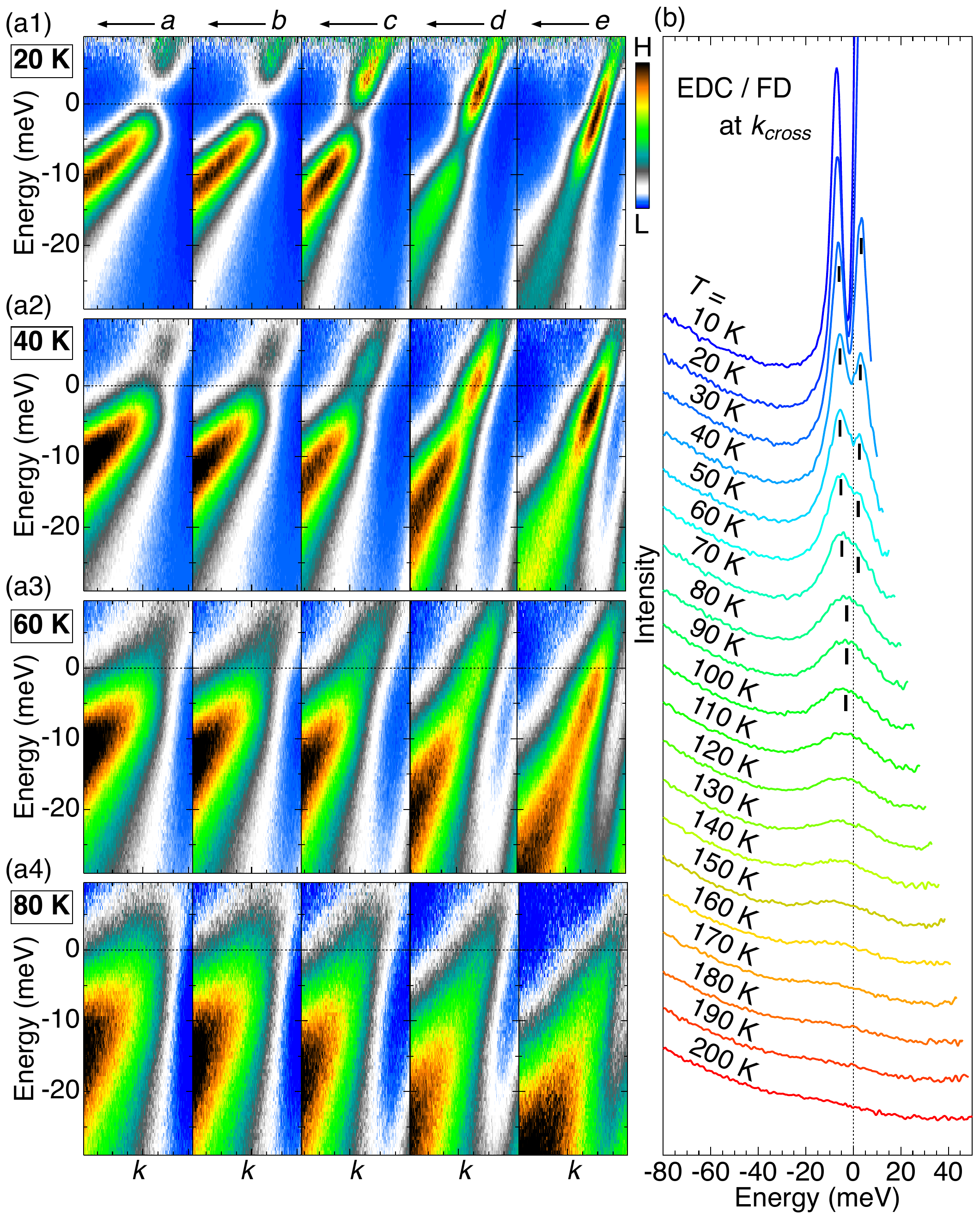}
\caption{Collapse of hybridization gap by coherent-incoherent crossover at elevated temperature.  (a1-a4) ARPES dispersions along several momentum cuts indicated by arrows in Fig.~ 3(a). Each image is divided by the measurement temperature:  20K (a1), 40 K (a2), 60 K (a3), and 80 K (a4). (b) EDCs at $k_{\rm cross}$ [see the inset of Fig.~3(b)]. The spectral peaks are marked by black bars. 
}  
\label{fig1}
\end{figure}

The effective SOC enhanced by strong electron correlations increases the orbital mixings. In Fig.~2(b), the calculated orbital weight of $d_{xy}$ and $d_{xz/yz}$ is represented by red and blue colors, respectively, for the FSs. Without SOC, no orbital mixing occurs, and each FS sheet is fully dominated by the $d_{xy}$ or $d_{xz/yz}$ component. Pronounced mixing occurs with SOC, and it is more prompted with increasing ${\lambda}$. 
This trend is represented in Fig.~2(b) with a more whitish color for the larger mixing ratio of $d_{xy}$ and $d_{xz/yz}$. For easy understanding, we illustrate the mixing ratio by a grayscale in Fig.~2(c), defining 100 $\%$ (black) when the wave function of $d_{xy}$ and $d_{xz/yz}$ is fully mixed; reversely, it is 0 $\%$ (white) when only one of the two orbitals ($d_{xy}$ and $d_{xz/yz}$) contributes. 
Plotting the mixing ratio around the pocket [Fig.~2(f)] clarifies the orbital selective nature seen at ${\lambda}$ = ${\lambda}_{\rm DFT}$ (100 meV) to be significantly reduced with increasing ${\lambda}$. At ${\lambda}$ = 250 meV we obtained, the mixing ratio becomes more than 50 $\%$ all around the Fermi pockets, lifting the orbital selectivity. 

The hybridization gap should also have a direct relation with SOC. In Fig.~2(d), we plot the calculated band dispersions along momenta going through the crossing point between $\beta'-$FS and $\gamma-$FS [a green dotted arrow in Fig.~2(b)] for different ${\lambda}$s. As summarized in Fig.~2(e), the hybridization gap (${\Delta}_{\rm calc}$) opens with a finite SOC, monotonically increasing against ${\lambda}$. The ${\Delta}_{\rm calc}$ at ${\lambda}$ = 250 meV is 66 meV, which is much larger than the experimental value (${\Delta}_{\rm ARPES}$ = $10\pm0.5$ meV). Note, however, that the experimental hybridization gap is renormalized by strong correlations~\cite{Kim2018hj}. The renormalization factor for the Fermi pockets (1/$Z_{\rm pocket}$$\equiv$$1/\sqrt{Z_{\beta}Z_{\gamma}}$) is estimated to be 6.0. Thus, the renormalized hybridization gap to be compared with the experimental value is obtained as ${Z_{\rm pocket}}{\Delta}_{\rm calc}$ = 11 meV, which well agrees with ${\Delta}_{\rm ARPES}$. Interestingly, this agreement is reached with physically different two effects of the electron correlation: band renormalization and SOC enhancement, which is induced by diagonal and off-diagonal orbital components of the self-energy. 

To examine the hybridization gap further, we plot in Fig.~3(b) the EDCs extracted at $k_{\rm cross}$ points (defined in the inset) marked by color circles in Fig.~3(a). We find that a definable spectral gap disappears below $\sim-10$ meV. This indicates that band hybridization is not well defined away from $E_{\rm F}$ most likely because the quasiparticle picture collapses. In Figs.~3(c2) and 3(d2), we examine the quasiparticle properties by plotting EDCs at several $k$ points marked by white arrows in Figs.~3(c1) and 3(d1) obtained along purple and orange dashed arrows in Fig.~3(a), respectively. Indeed, the quasiparticle peaks disappear when the bands go off $E_{\rm F}$ and get incoherent. Intriguingly, the energy window where the quasiparticles survive is extremely small with only $\sim$20 meV about $E_{\rm F}$. It is a much smaller energy scale than for the bulk state of Sr$_2$RuO$_4$ ($\sim40$~meV~\cite{Damascelli_FS}) and strongly correlated $3d$-orbital systems as cuprates ($\sim70$~meV). 

The coherent-incoherent crossover with temperature is also an intrinsic property of Hund’s metal. In Fig.~4(a1), we plot the FD-divided ARPES dispersions at 20 K for several momentum cuts [black arrows in Fig.~3(a)]. 
Similar data sets are also taken at higher temperatures 40 K, 60 K, and 80 K in Fig.~4(a2), 4(a3), and 4(a4), respectively. The disconnected hybridization gap gets unclear with increasing temperature, and eventually disappears around 80 K, having the band a continuous dispersion of broad spectra [Fig.~4(a4)]. 
More detail is examined in Fig.~4(b) by plotting the temperature dependence of EDCs at ${k_{\rm cross}}$ for the momentum cut $c$ [described on panel (a1)]. With increasing temperature, two peaks indicative of a band gap are significantly suppressed and broadened, and the spectral line shape eventually becomes a broad one-peak structure with the spectral gap closed around 80 K. 
This indicates that band hybridization cannot be well-defined quantum mechanically anymore due to the collapsing of the quasiparticle picture above $\sim$80~K. With further increasing temperatures, the spectra become completely incoherent with an almost flat shape. We confirmed that this behavior is not due to the aging of the sample surface (see Supplemental Fig. S1~\cite{SM}). We also note that the rotation angle of RuO$_6$ does not change with temperature~\cite{STM,Peter_Advanced,Moore_SurfacePhase}. 
According to theory, the coherent-incoherent crossover is not reproduced only with the on-site Coulomb interaction $U$ (Mott physics) but requires Hund's coupling $J$ (Hund physics)~\cite{HundMetal1,HundMetal_PRB}. Hence, the dramatic variations in the hybridized spectra we observe should reflect a specific feature of Hund's metal state. 

In conclusion, we revealed anomalous features of Fermi pockets on the Sr$_2$RuO$_4$ surface. The Fermi pockets were found to be significantly deformed by correlation-enhanced SOC. The enhancement of SOC is much larger than that of the bulk state, causing high mixing of $d_{xy}$ and  $d_{xz/yz}$ orbitals. This would be the reason that the orbital-selective Mott transition (OSMT) is hard to be realized in the perovskite oxides with octahedral rotation. 
In particular, although Ca$_{2-x}$Sr$_x$RuO$_4$ ($0.2 \le x \le 0.5$) is the most debated OSMT candidate~\cite{AnisimovEPB,Liebsch_2003,LiebschPRL2003,LiebschPRB,KogaPRL,LiebschPRL,BalicasPRL,LeePRL,StepanovPRL}, experiments reveal that their bands ($\alpha$-, $\beta$-, and $\gamma$ bands) are all metallic with no gap~\cite{WangYangPRL,ShimoyamadaPRL}, thus being negative on this proposal. Ca$_{2-x}$Sr$_x$RuO$_4$ has the rotational angle even greater ($\sim 12^{\circ}$) than that of the Sr$_2$RuO$_4$ surface ($\sim 8^{\circ}$) and it is also affected by the octahedral tilt~\cite{CadopeRotation}, which should further enhance the effective SOC. 
Our results, therefore, reasonably explain why OSMT is not realized; that is, the effective SOC is so much enhanced by crystal distortion (rotation and tilt) that the orbital selectivity is lifted in these systems. 

Here, we point out that the previous DMFT calculations reproduced experimental results of Sr$_2$RuO$_4$ in which band renormalization is greatly enhanced in the surface band compared to the bulk band, only by including the effect of octahedral rotation on hopping integrals~\cite{KondoPRL}. This indicates that Hund's coupling $J$ is comparable between the bulk and the surface, supporting our argument that the enhanced SOC due to crystal distortion precludes OSMT by outweighing the effect of $J$ that causes orbital isolation, in a system such as Ca$_{2-x}$Sr$_x$RuO$_4$. 

We also found a hybridization gap to be enlarged by the correlation-enhanced SOC. Moreover, Hund’s metal features restrict the well-defined band hybridization to energies extremely close to $E_{\rm F}$ and eliminate the hybridization gap during the coherent-incoherent crossover at elevated temperatures. These anomalies of the Fermi pockets are featured by dichotomic properties of orbital-isolating Hund’s coupling and orbital-mixing SOC which are both related with intriguing strong correlation effects.
These notions apply not only to the Sr$_2$RuO$_4$ surface but, more generally, to many other perovskite oxides which are commonly accompanied by octahedral rotation.

This work was supported by the JSPS KAKENHI (Grant No. JP21H04439 and JP23K17351), by the Asahi Glass Foundation, by The Murata Science Foundation, and by MEXT Q-LEAP (Grant No. JPMXS0118068681). 



\begin{thebibliography}{10}

\bibitem{Borisenko}
S.~V. Borisenko {\it et~al.}
\newblock Direct observation of spin--orbit coupling in iron-based
  superconductors.
\newblock {\em Nat. Phys.}{ \bf 12}, 311--317 (2016).

\bibitem{Balents}
W.~Witczak-Krempa, G.~Chen, Y.~B. Kim, and L.~Balents.
\newblock Correlated Quantum Phenomena in the Strong Spin-Orbit Regime.
\newblock {\em Annual Review of Condensed Matter Physics}{ \bf 5}, 57--82
  (2014).

\bibitem{Damascelli_IronSC}
R.~P. Day {\it et~al.}
\newblock Influence of Spin-Orbit Coupling in Iron-Based Superconductors.
\newblock {\em Phys. Rev. Lett.}{ \bf 121}, 076401 (2018).

\bibitem{Taylor}
A.~E. Taylor, S.~Calder, R.~Morrow, H.~L. Feng, M.~H. Upton, M.~D. Lumsden,
  K.~Yamaura, P.~M. Woodward, and A.~D. Christianson.
\newblock Spin-Orbit Coupling Controlled $J=3/2$ Electronic Ground State in
  $5{d}^{3}$ Oxides.
\newblock {\em Phys. Rev. Lett.}{ \bf 118}, 207202 (2017).

\bibitem{Johnson}
P.~D. Johnson, H.-B. Yang, J.~D. Rameau, G.~D. Gu, Z.-H. Pan, T.~Valla,
  M.~Weinert, and A.~V. Fedorov.
\newblock Spin-Orbit Interactions and the Nematicity Observed in the Fe-Based
  Superconductors.
\newblock {\em Phys. Rev. Lett.}{ \bf 114}, 167001 (2015).

\bibitem{Stadler}
H.~Miao {\it et~al.}
\newblock Dynamical Mean-Field Theory Plus Numerical Renormalization-Group
  Study of Spin-Orbital Separation in a Three-Band Hund Metal.
\newblock {\em Phys. Rev. Lett.}{ \bf 115}, 136401 (2015).

\bibitem{Day}
R.~P. Day {\it et~al.}
\newblock Influence of Spin-Orbit Coupling in Iron-Based Superconductors.
\newblock {\em Phys. Rev. Lett.}{ \bf 121}, 076401 (2018).

\bibitem{Vale}
J.~G. Vale {\it et~al.}
\newblock Spin-orbit-driven magnetic structure and excitation in the 5d
  pyrochlore Cd$_2$Os$_2$O$_7$.
\newblock {\em Nat. Commun.}{ \bf 7}, 1--8 (2016).

\bibitem{HiroshiShinaoka}
H.~Shinaoka, T.~Miyake, and S.~Ishibashi.
\newblock Noncollinear Magnetism and Spin-Orbit Coupling in $5d$ Pyrochlore
  Oxide ${\mathrm{Cd}}_{2}{\mathrm{Os}}_{2}{\mathrm{O}}_{7}$.
\newblock {\em Phys. Rev. Lett.}{ \bf 108}, 247204 (2012).

\bibitem{Kim2018hj}
M.~Kim, J.~Mravlje, M.~Ferrero, O.~Parcollet, and A.~Georges.
\newblock Spin-Orbit Coupling and Electronic Correlations in Sr$_{2}$RuO$_{4}$.
\newblock {\em Phys. Rev. Lett.}{ \bf 120}, 126401 (2018).

\bibitem{Arita}
R.~Arita, J.~Kune\ifmmode~\check{s}\else \v{s}\fi{}, A.~V. Kozhevnikov, A.~G.
  Eguiluz, and M.~Imada.
\newblock Ab initio Studies on the Interplay between Spin-Orbit Interaction and
  Coulomb Correlation in ${\mathrm{Sr}}_{2}{\mathrm{IrO}}_{4}$ and
  ${\mathrm{Ba}}_{2}{\mathrm{IrO}}_{4}$.
\newblock {\em Phys. Rev. Lett.}{ \bf 108}, 086403 (2012).

\bibitem{Zhang2016bz}
G.~Zhang, E.~Gorelov, E.~Sarvestani, and E.~Pavarini.
\newblock Fermi Surface of Sr$_{2}$RuO$_{4}$: Spin-Orbit and Anisotropic
  Coulomb Interaction Effects.
\newblock {\em Phys. Rev. Lett.}{ \bf 116}, 106402 (2016).

\bibitem{Tamai2019ef}
A.~Tamai {\it et~al.}
\newblock High-Resolution Photoemission on Sr$_{2}$RuO$_{4}$ Reveals
  Correlation-Enhanced Effective Spin-Orbit Coupling and Dominantly Local
  Self-Energies.
\newblock {\em Phys. Rev. X}{ \bf 9}, 021048 (2019).

\bibitem{Damascelli2014}
C.~N. Veenstra {\it et~al.}
\newblock Spin-Orbital Entanglement and the Breakdown of Singlets and Triplets
  in ${\mathrm{Sr}}_{2}{\mathrm{RuO}}_{4}$ Revealed by Spin- and Angle-Resolved
  Photoemission Spectroscopy.
\newblock {\em Phys. Rev. Lett.}{ \bf 112}, 127002 (2014).

\bibitem{Iwasawa}
H.~Iwasawa, Y.~Yoshida, I.~Hase, S.~Koikegami, H.~Hayashi, J.~Jiang,
  K.~Shimada, H.~Namatame, M.~Taniguchi, and Y.~Aiura.
\newblock Interplay among Coulomb Interaction, Spin-Orbit Interaction, and
  Multiple Electron-Boson Interactions in Sr$_{2}$RuO$_{4}$.
\newblock {\em Phys. Rev. Lett.}{ \bf 105}, 123702 (2010).

\bibitem{Haverkort}
M.~W. Haverkort, I.~S. Elfimov, L.~H. Tjeng, G.~A. Sawatzky, and A.~Damascelli.
\newblock Strong Spin-Orbit Coupling Effects on the Fermi Surface of
  ${\mathrm{Sr}}_{2}{\mathrm{RuO}}_{4}$ and
  ${\mathrm{Sr}}_{2}{\mathrm{RhO}}_{4}$.
\newblock {\em Phys. Rev. Lett.}{ \bf 101}, 026406 (2008).

\bibitem{DJKim_PRL}
B.~Kim {\it et~al.}
\newblock {Novel Jeff=1/2 Mott State Induced by Relativistic Spin-Orbit
  Coupling in Sr$_{2}$IrO$_{4}$}.
\newblock {\em Phys. Rev. Lett.}{ \bf 101}, 076402 (2008).

\bibitem{LukeJSandilands}
L.~J. Sandilands, W.~Kyung, S.~Y. Kim, J.~Son, J.~Kwon, T.~D. Kang, Y.~Yoshida,
  S.~J. Moon, C.~Kim, and T.~W. Noh.
\newblock Spin-Orbit Coupling and Interband Transitions in the Optical
  Conductivity of ${\mathrm{Sr}}_{2}{\mathrm{RhO}}_{4}$.
\newblock {\em Phys. Rev. Lett.}{ \bf 119}, 267402 (2017).

\bibitem{Zwartsenberg}
B.~Zwartsenberg {\it et~al.}
\newblock Spin-orbit-controlled metal--insulator transition in
  Sr$_{2}$IrO$_{4}$.
\newblock {\em Nat. Phys.}{ \bf 16}, 290--294 (2020).

\bibitem{KimAaram}
A.~J. Kim, H.~O. Jeschke, P.~Werner, and R.~Valent\'{\i}.
\newblock $\mathbf{J}$ Freezing and Hund's Rules in Spin-Orbit-Coupled
  Multiorbital Hubbard Models.
\newblock {\em Phys. Rev. Lett.}{ \bf 118}, 086401 (2017).

\bibitem{Dhital}
C.~Dhital {\it et~al.}
\newblock Carrier localization and electronic phase separation in a doped
  spin-orbit-driven Mott phase in Sr$_{3}$(Ir$_{1-x}$Ru$_x$)$_2$O$_{7}$.
\newblock {\em Nat. Commun.}{ \bf 5}, 1--7 (2014).

\bibitem{Das}
L.~Das {\it et~al.}
\newblock Spin-Orbital Excitations in Ca$_{2}$RuO$_{4}$ Revealed by Resonant
  Inelastic X-Ray Scattering.
\newblock {\em Phys. Rev. X}{ \bf 8}, 011048 (2018).

\bibitem{ZhouSen}
S.~Zhou, K.~Jiang, H.~Chen, and Z.~Wang.
\newblock Correlation Effects and Hidden Spin-Orbit Entangled Electronic Order
  in Parent and Electron-Doped Iridates Sr$_{2}$IrO$_{4}$.
\newblock {\em Phys. Rev. X}{ \bf 7}, 041018 (2017).

\bibitem{Cui}
Q.~Cui {\it et~al.}
\newblock Slater Insulator in Iridate Perovskites with Strong Spin-Orbit
  Coupling.
\newblock {\em Phys. Rev. Lett.}{ \bf 117}, 176603 (2016).

\bibitem{Watanabe}
H.~Watanabe, T.~Shirakawa, and S.~Yunoki.
\newblock Microscopic Study of a Spin-Orbit-Induced Mott Insulator in Ir
  Oxides.
\newblock {\em Phys. Rev. Lett.}{ \bf 105}, 216410 (2010).

\bibitem{Nie}
Y.~F. Nie {\it et~al.}
\newblock Interplay of Spin-Orbit Interactions, Dimensionality, and Octahedral
  Rotations in Semimetallic ${\mathbf{\text{SrIrO}}}_{3}$.
\newblock {\em Phys. Rev. Lett.}{ \bf 114}, 016401 (2015).

\bibitem{Martins}
C.~Martins, M.~Aichhorn, L.~Vaugier, and S.~Biermann.
\newblock Reduced Effective Spin-Orbital Degeneracy and Spin-Orbital Ordering
  in Paramagnetic Transition-Metal Oxides:
  ${\mathrm{Sr}}_{2}{\mathrm{IrO}}_{4}$ versus
  ${\mathrm{Sr}}_{2}{\mathrm{RhO}}_{4}$.
\newblock {\em Phys. Rev. Lett.}{ \bf 107}, 266404 (2011).

\bibitem{Rotation_PRB}
R.~Matzdorf, Ismail, T.~Kimura, Y.~Tokura, and E.~W. Plummer.
\newblock Surface structural analysis of the layered perovskite
  ${\mathrm{Sr}}_{2}{\mathrm{RuO}}_{4}$ by LEED $I(V)$.
\newblock {\em Phys. Rev. B}{ \bf 65}, 085404 (2002).

\bibitem{Sr3Ir2O7_rotation}
M.~Subramanian, M.~Crawford, and R.~Harlow.
\newblock Single crystal structure determination of double layered strontium
  iridium oxide [Sr$_3$Ir$_2$O$_7$].
\newblock {\em Mater. Res. Bull.}{ \bf 29}, 645--650 (1994).

\bibitem{Sr3Ru2O7_rotation}
B.~Hu, G.~T. McCandless, M.~Menard, V.~B. Nascimento, J.~Y. Chan, E.~W.
  Plummer, and R.~Jin.
\newblock Surface and bulk structural properties of single-crystalline
  ${\text{Sr}}_{3}{\text{Ru}}_{2}{\text{O}}_{7}$.
\newblock {\em Phys. Rev. B}{ \bf 81}, 184104 (2010).

\bibitem{Sr2IrO4_rotation}
M.~K. Crawford, M.~A. Subramanian, R.~L. Harlow, J.~A. Fernandez-Baca, Z.~R.
  Wang, and D.~C. Johnston.
\newblock Structural and magnetic studies of
  ${\mathrm{Sr}}_{2}$${\mathrm{IrO}}_{4}$.
\newblock {\em Phys. Rev. B}{ \bf 49}, 9198--9201 (1994).

\bibitem{CadopeRotation}
O.~Friedt, M.~Braden, G.~Andr\'e, P.~Adelmann, S.~Nakatsuji, and Y.~Maeno.
\newblock Structural and magnetic aspects of the metal-insulator transition in
  ${\mathrm{Ca}}_{2\ensuremath{-}x}{\mathrm{Sr}}_{x}{\mathrm{RuO}}_{4}$.
\newblock {\em Phys. Rev. B}{ \bf 63}, 174432 (2001).

\bibitem{Braden4}
M.~Braden, W.~Reichardt, S.~Nishizaki, Y.~Mori, and Y.~Maeno.
\newblock Structural stability of Sr$_2$RuO$_4$.
\newblock {\em Phys. Rev. B}{ \bf 57}, 1236 (1998).

\bibitem{Moore_SurfacePhase}
R.~G. Moore, V.~B. Nascimento, J.~Zhang, J.~Rundgren, R.~Jin, D.~Mandrus, and
  E.~W. Plummer.
\newblock Manifestations of Broken Symmetry: The Surface Phases of
  ${\mathrm{Ca}}_{2\ensuremath{-}x}{\mathrm{Sr}}_{x}{\mathrm{RuO}}_{4}$.
\newblock {\em Phys. Rev. Lett.}{ \bf 100}, 066102 (2008).

\bibitem{STM}
R.~Matzdorf, Z.~Fang, J.~Zhang, T.~Kimura, Y.~Tokura, K.~Terakura, and E.~W.
  Plummer.
\newblock {Ferromagnetism stabilized by lattice distortion at the surface of
  the p-wave superconductor Sr$_2$RuO$_4$}.
\newblock {\em Science}{ \bf 289}, 746--748 (2000).

\bibitem{Ko_DFT}
E.~Ko, B.~Kim, C.~Kim, and H.~Choi.
\newblock {Strong Orbital-Dependent d-Band Hybridization and Fermi-Surface
  Reconstruction in Metallic Ca$_{2-x}$Sr$_x$RuO$_4$}.
\newblock {\em Phys. Rev. Lett.}{ \bf 98}, 226401 (2007).

\bibitem{AnisimovEPB}
V.~Anisimov, I.~Nekrasov, D.~Kondakov, T.~Rice, and M.~Sigrist.
\newblock Orbital-selective Mott-insulator transition in
  Ca$_{2-x}$Sr$_x$RuO$_4$.
\newblock {\em Eur Phys. J. B}{ \bf 25}, 191--201 (2002).

\bibitem{Liebsch_2003}
A.~Liebsch.
\newblock Absence of orbital-dependent Mott transition in
  Ca$_{2-x}$Sr$_x$RuO$_4$.
\newblock {\em Europhys. Lett.}{ \bf 63}, 97 (2003).

\bibitem{LiebschPRL2003}
A.~Liebsch.
\newblock Mott transitions in multiorbital systems.
\newblock {\em Phys. Rev. Lett.}{ \bf 91}, 226401 (2003).

\bibitem{LiebschPRB}
A.~Liebsch.
\newblock Single Mott transition in the multiorbital Hubbard model.
\newblock {\em Phys. Rev. B}{ \bf 70} (2004).

\bibitem{WangYangPRL}
S.-C. Wang {\it et~al.}
\newblock Fermi surface topology of Ca$_{1.5}$Sr$_{0.5}$RuO$_4$ determined by
  angle-resolved photoelectron spectroscopy.
\newblock {\em Phys. Rev. Lett.}{ \bf 93}, 177007 (2004).

\bibitem{ShimoyamadaPRL}
A.~Shimoyamada, K.~Ishizaka, S.~Tsuda, S.~Nakatsuji, Y.~Maeno, and S.~Shin.
\newblock Strong mass renormalization at a local momentum space in multiorbital
  Ca$_{1.8}$Sr$_{0.2}$RuO$_4$.
\newblock {\em Phys. Rev. Lett.}{ \bf 102}, 086401 (2009).

\bibitem{Ir_rotation}
M.~K. Crawford, M.~A. Subramanian, R.~L. Harlow, J.~A. Fernandez-Baca, Z.~R.
  Wang, and D.~C. Johnston.
\newblock Structural and magnetic studies of
  ${\mathrm{Sr}}_{2}$${\mathrm{IrO}}_{4}$.
\newblock {\em Phys. Rev. B}{ \bf 49}, 9198--9201 (1994).

\bibitem{Kim_arc1}
Y.~Kim, O.~Krupin, J.~Denlinger, A.~Bostwick, E.~Rotenberg, Q.~Zhao,
  J.~Mitchell, J.~Allen, and B.~Kim.
\newblock Fermi arcs in a doped pseudospin-1/2 Heisenberg antiferromagnet.
\newblock {\em Science}{ \bf 345}, 187--190 (2014).

\bibitem{Kim_arc2}
Y.~K. Kim, N.~H. Sung, J.~D. Denlinger, and B.~J. Kim.
\newblock {Observation of a d-wave gap in electron-doped Sr$_{2}$IrO$_{4}$}.
\newblock {\em Nat. Phys.}{ \bf 12}, 37--41 (2015).

\bibitem{Sr2IrO4_Baumberger}
A.~de~la Torre {\it et~al.}
\newblock {Collapse of the Mott Gap and Emergence of a Nodal Liquid in Lightly
  Doped Sr$_{2}$IrO$_{4}$}.
\newblock {\em Phys. Rev. Lett.}{ \bf 115}, 176402 (2015).

\bibitem{Sr2RhO4_Kim}
B.~Kim, J.~Yu, H.~Koh, I.~Nagai, S.~Ikeda, S.-J. Oh, and C.~Kim.
\newblock {Missing xy-Band Fermi Surface in 4d Transition-Metal Oxide
  Sr$_{2}$RhO$_{4}$: Effect of the Octahedra Rotation on the Electronic
  Structure}.
\newblock {\em Phys. Rev. Lett.}{ \bf 97}, 106401 (2006).

\bibitem{Sr2RhO4_Baumberger}
F.~Baumberger, N.~Ingle, W.~Meevasana, K.~Shen, D.~Lu, R.~Perry, A.~Mackenzie,
  Z.~Hussain, D.~Singh, and Z.-X. Shen.
\newblock {Fermi Surface and Quasiparticle Excitations of Sr$_{2}$RhO$_{4}$}.
\newblock {\em Phys. Rev. Lett.}{ \bf 96}, 246402 (2006).

\bibitem{Damascelli_FS}
A.~Damascelli {\it et~al.}
\newblock {Fermi Surface, Surface States, and Surface Reconstruction in
  Sr$_2$RuO$_4$}.
\newblock {\em Phys. Rev. Lett.}{ \bf 85}, 5194--5197 (2000).

\bibitem{Shen_surface}
K.~Shen {\it et~al.}
\newblock {Surface electronic structure of Sr$_2$RuO$_4$}.
\newblock {\em Phys. Rev. B}{ \bf 64}, 180502 (2001).

\bibitem{Damascelli_Progression}
C.~N. Veenstra {\it et~al.}
\newblock Determining the Surface-To-Bulk Progression in the Normal-State
  Electronic Structure of ${\mathrm{Sr}}_{2}{\mathrm{RuO}}_{4}$ by
  Angle-Resolved Photoemission and Density Functional Theory.
\newblock {\em Phys. Rev. Lett.}{ \bf 110}, 097004 (2013).

\bibitem{KondoPRL}
T.~Kondo {\it et~al.}
\newblock Orbital-Dependent Band Narrowing Revealed in an Extremely Correlated
  Hund's Metal Emerging on the Topmost Layer of
  ${\mathrm{Sr}}_{2}{\mathrm{RuO}}_{4}$.
\newblock {\em Phys. Rev. Lett.}{ \bf 117}, 247001 (2016).

\bibitem{HundMetal1}
K.~M. Stadler, Z.~P. Yin, J.~von Delft, G.~Kotliar, and A.~Weichselbaum.
\newblock {Dynamical Mean-Field Theory Plus Numerical Renormalization-Group
  Study of Spin-Orbital Separation in a Three-Band Hund Metal}.
\newblock {\em Phys. Rev. Lett.}{ \bf 115}, 136401 (2015).

\bibitem{HundMetal_NatNano}
A.~Khajetoorians, M.~Valentyuk, M.~Steinbrecher, T.~Schlenk, A.~Shick,
  J.~Kolorenc, A.~Lichtenstein, T.~Wehling, R.~Wiesendanger, and J.~Wiebe.
\newblock {Tuning emergent magnetism in a Hund's impurity}.
\newblock {\em Nat. Nanotechnol.}{ \bf 10}, 958 --964 (2015).

\bibitem{HundMetal_resisitivity}
X.~Deng, K.~Haule, and G.~Kotliar.
\newblock {Transport Properties of Metallic Ruthenates: A
  DFT+DMFTInvestigation}.
\newblock {\em Phys. Rev. Lett.}{ \bf 116}, 256401 (2016).

\bibitem{Hund_Seebeck}
J.~Mravlje and A.~Georges.
\newblock {Thermopower and Entropy: Lessons from Sr$_{2}$RuO$_{4}$}.
\newblock {\em Phys. Rev. Lett.}{ \bf 117}, 036401 (2016).

\bibitem{HundMetal2}
L.~De'~Medici, J.~Mravlje, and A.~Georges.
\newblock {Janus-Faced Influence of Hund's Rule Coupling in Strongly Correlated
  Materials}.
\newblock {\em Phys. Rev. Lett.}{ \bf 107}, 256401 (2011).

\bibitem{Crossover}
J.~Mravlje, M.~Aichhorn, T.~Miyake, K.~Haule, G.~Kotliar, and A.~Georges.
\newblock Coherence-Incoherence Crossover and the Mass-Renormalization Puzzles
  in ${\mathrm{Sr}}_{2}{\mathrm{RuO}}_{4}$.
\newblock {\em Phys. Rev. Lett.}{ \bf 106}, 096401 (2011).

\bibitem{HundMetal_PRB}
L.~Fanfarillo and E.~Bascones.
\newblock {Electronic correlations in Hund metals}.
\newblock {\em Phys. Rev. B}{ \bf 92}, 075136 (2015).

\bibitem{MAO20001813}
Z.~Mao, Y.~Maenoab, and H.~Fukazawa.
\newblock Crystal growth of Sr$_{2}$RuO$_{4}$.
\newblock {\em Mater. Res. Bull.}{ \bf 35}, 1813--1824 (2000).

\bibitem{SM}
See Supplemental Material, which includes Refs. [70-79], for aging check of
  sample surface after temperature scan measurements; momentum dependence of
  the hybridization gap; and details of our first-principles DFT band
  calculations.

\bibitem{KingSurfacePRL}
E.~Abarca~Morales {\it et~al.}
\newblock Hierarchy of Lifshitz Transitions in the Surface Electronic Structure
  of ${\mathrm{Sr}}_{2}{\mathrm{RuO}}_{4}$ under Uniaxial Compression.
\newblock {\em Phys. Rev. Lett.}{ \bf 130}, 096401 (2023).

\bibitem{Peter_Advanced}
C.~A. Marques, L.~C. Rhodes, R.~Fittipaldi, V.~Granata, C.~M. Yim, R.~Buzio,
  A.~Gerbi, A.~Vecchione, A.~W. Rost, and P.~Wahl.
\newblock Magnetic-Field Tunable Intertwined Checkerboard Charge Order and
  Nematicity in the Surface Layer of Sr$_2$RuO$_4$.
\newblock {\em Advanced materials (Deerfield Beach, Fla.)}{ \bf 33}, e2100593
  (2021).

\bibitem{Maeno}
A.~P. Mackenzie and Y.~Maeno.
\newblock The superconductivity of Sr$_{2}$RuO$_{4}$ and the physics of
  spin-triplet pairing.
\newblock {\em Rev. Mod. Phys.}{ \bf 75}, 657 (2003).

\bibitem{KogaPRL}
A.~Koga, N.~Kawakami, T.~M. Rice, and M.~Sigrist.
\newblock Orbital-selective mott transitions in the degenerate Hubbard model.
\newblock {\em Phys. Rev. Lett.}{ \bf 92}, 216402 (2004).

\bibitem{LiebschPRL}
A.~Liebsch.
\newblock Novel Mott transitions in a nonisotropic two-band Hubbard model.
\newblock {\em Phys. Rev. Lett.}{ \bf 95}, 116402 (2005).

\bibitem{BalicasPRL}
L.~Balicas, S.~Nakatsuji, D.~Hall, T.~Ohnishi, Z.~Fisk, Y.~Maeno, and D.~J.
  Singh.
\newblock Severe Fermi Surface Reconstruction at a Metamagnetic Transition in
  ${\mathrm{Ca}}_{2\ensuremath{-}x}{\mathrm{Sr}}_{x}{\mathrm{RuO}}_{4}$ (for
  $0.2\ensuremath{\le}x\ensuremath{\le}0.5$).
\newblock {\em Phys. Rev. Lett.}{ \bf 95}, 196407 (2005).

\bibitem{LeePRL}
J.~S. Lee, S.~J. Moon, T.~W. Noh, S.~Nakatsuji, and Y.~Maeno.
\newblock Orbital-Selective Mass Enhancements in Multiband
  ${\mathrm{Ca}}_{2\ensuremath{-}x}{\mathrm{Sr}}_{x}{\mathrm{RuO}}_{4}$ Systems
  Analyzed by the Extended Drude Model.
\newblock {\em Phys. Rev. Lett.}{ \bf 96}, 057401 (2006).

\bibitem{StepanovPRL}
E.~A. Stepanov.
\newblock Eliminating Orbital Selectivity from the Metal-Insulator Transition
  by Strong Magnetic Fluctuations.
\newblock {\em Phys. Rev. Lett.}{ \bf 129}, 096404 (2022).

\bibitem{PBE}
J.~P. Perdew, K.~Burke, and M.~Ernzerhof.
\newblock Generalized Gradient Approximation Made Simple.
\newblock {\em Phys. Rev. Lett.}{ \bf 77}, 3865--3868 (1996).

\bibitem{paw}
G.~Kresse and D.~Joubert.
\newblock From ultrasoft pseudopotentials to the projector augmented-wave
  method.
\newblock {\em Phys. Rev. B}{ \bf 59}, 1758--1775 (1999).

\bibitem{vasp1}
G.~Kresse and J.~Hafner.
\newblock Ab initio molecular dynamics for liquid metals.
\newblock {\em Phys. Rev. B}{ \bf 47}, 558--561 (1993).

\bibitem{vasp2}
G.~Kresse and J.~Hafner.
\newblock Ab initio molecular-dynamics simulation of the
  liquid-metal--amorphous-semiconductor transition in germanium.
\newblock {\em Phys. Rev. B}{ \bf 49}, 14251--14269 (1994).

\bibitem{vasp3}
G.~Kresse and J.~Furthm\"{u}ller.
\newblock Efficiency of ab-initio total energy calculations for metals and
  semiconductors using a plane-wave basis set.
\newblock {\em Computational Materials Science}{ \bf 6}, 15--50 (1996).

\bibitem{vasp4}
G.~Kresse and J.~Furthm\"{u}ller.
\newblock Efficient iterative schemes for ab initio total-energy calculations
  using a plane-wave basis set.
\newblock {\em Phys. Rev. B}{ \bf 54}, 11169--11186 (1996).

\bibitem{struct2}
Q.~Huang, J.~Soubeyroux, O.~Chmaissem, I.~Sora, A.~Santoro, R.~Cava,
  J.~Krajewski, and W.~Peck.
\newblock Neutron Powder Diffraction Study of the Crystal Structures of Sr2RuO4
  and Sr2IrO4 at Room Temperature and at 10 K.
\newblock {\em Journal of Solid State Chemistry}{ \bf 112}, 355--361 (1994).

\bibitem{Wannier1}
N.~Marzari and D.~Vanderbilt.
\newblock Maximally localized generalized Wannier functions for composite
  energy bands.
\newblock {\em Phys. Rev. B}{ \bf 56}, 12847--12865 (1997).

\bibitem{Wannier2}
I.~Souza, N.~Marzari, and D.~Vanderbilt.
\newblock Maximally localized Wannier functions for entangled energy bands.
\newblock {\em Phys. Rev. B}{ \bf 65}, 035109 (2001).

\bibitem{Wannier90}
G.~Pizzi {\it et~al.}
\newblock Wannier90 as a community code: new features and applications.
\newblock {\em Journal of Physics: Condensed Matter}{ \bf 32}, 165902 (2020).

\end{thebibliography}

\end{document}